\documentstyle[prd,aps,epsfig,floats]{revtex}
\tighten

\newcommand{\al}{\alpha}

\newcommand{\de}{\delta}

\newcommand{\ep}{\epsilon}

\newcommand{\La}{\Lambda}
\newcommand{\la}{\lambda}
\newcommand{\Om}{\Omega}

\newcommand{\si}{\sigma}
\newcommand{\Si}{\Sigma}

\newcommand{\Th}{\Theta}

\newcommand{\ra}{\rightarrow}
\newcommand{\be}{\begin{equation}}
\newcommand{\ee}{\end{equation}}
\newcommand{\bea}{\begin{eqnarray}}
\newcommand{\eea}{\end{eqnarray}}
\newcommand{\bean}{\begin{eqnarray*}}
\newcommand{\eean}{\end{eqnarray*}}

\def\lsim{\raise 0.4ex\hbox{$<$}\kern -0.8em\lower 0.62ex\hbox{$\sim$}}
\def\gsim{\raise 0.4ex\hbox{$>$}\kern -0.7em\lower 0.62ex\hbox{$\sim$}}
\newcommand{\bk}{{\bf k}}

\begin{document}
\draft
\twocolumn[\hsize\textwidth\columnwidth\hsize\csname@twocolumnfalse\endcsname
\title{ Reproducing the observed Cosmic microwave background anisotropies
 with \\ causal scaling seeds}
\author{R. Durrer$^1$, M. Kunz$^2$ \and A. Melchiorri$^2$}
\address{$^1$D\'epartement de Physique Th\'eorique,
	 Universit\'e de Gen\`eve,
	24 quai Ernest Ansermet, CH-1211 Gen\`eve 4, Switzerland\\
        $^2$Astrophysics, Oxford University, Keble Road, Oxford OX1 3RH, UK}

\maketitle

\begin{abstract}
\begin{center}{\large\bf Abstract}\end{center} 
During the last years it has become clear that global $O(N)$
defects and $U(1)$ cosmic strings do not lead to the pronounced first
acoustic peak in the  power spectrum of anisotropies of the  cosmic
microwave background which has recently been observed  to high accuracy.
Inflationary models cannot easily accommodate the low second
peak indicated by the data. Here we construct causal scaling seed
models which reproduce the first and second peak.
Future, more precise CMB anisotropy and polarization experiments
will however be able to distinguish them from the ordinary
adiabatic models.
\end{abstract}
\date{today}
]
{
\section{Introduction}
Inflation and topological defects are  
two classes of models to explain the origin of large scale structure in 
the universe. In inflationary models, for fixed cosmological
parameters the fluctuation spectrum is determined by the initial
conditions. In models with topological defects or other types
of seeds, fluctuations in the cosmic plasma and in the geometry 
are continuously induced by the gravitational coupling to the seed
energy momentum tensor.

Cosmic microwave background~(CMB) anisotropies provide a excellent link 
between theoretical predictions and observational data. They allow us
to distinguish between inflationary perturbations and models with
defects by purely linear analysis.  On large angular scales, 
both classes of models 
predict an approximately scale-invariant Harrison-Zel'dovich spectrum
\cite{h,z}. For inflationary models this can be seen analytically. 
Scale-invariance for defects was discovered numerically
\cite{SPT,DZ,Shellard}; simple analytical arguments are given  {\em e.g.}
in \cite{Roma}.

On smaller angular scales ($10' \lsim \theta \lsim 2^{\circ}$), 
the predictions of inflation and global $O(N)$ defects
are different. While inflationary models predict
a series of 'acoustic peaks', global  $O(N)$ defects show a low
amplitude broad 'hump'~\cite{DGS,Seljak,DKM}. For local $U(1)$ cosmic
strings, the result is not
 so clear. Depending on the detailed modeling of local cosmic strings, 
the resulting acoustic peaks are quite
different. The peak can be entirely absent~\cite{Shellard} or present and
even quite substantial, but at an angular harmonic $\ell \sim 400$ -- 
$500$~~\cite{Albrecht,Joao,Vachaspati}.

Recent experiments~\cite{toco,B97,B98,MAX} have measured  CMB anisotropies 
which are fully compatible with a flat adiabatic inflationary model on the
scale of the first peak and incompatible with the above mentioned
defect models. However, the second peak is too low for
values of the baryon density that are within the constraints inferred
from standard nucleosynthesis. Combining the recent BOOMERanG
and MAXIMA-I data with informations from the distribution of galaxies, 
a value of $\Omega_bh^2=0.032\pm0.004$ was found in \cite{jaffe}
(see also \cite{pierpol,tegman,tegman2,lange}),
which is incompatible at nearly $3 \sigma$ with
the value $\Omega_bh^2=0.0189 \pm 0.0019$
 \cite{burles} inferred mainly from measurements of
primordial deuterium from Ly-alpha absorption systems
in the continuum emission of $3$ high redshift quasars.  

Even if it is fair to say that the possibility of systematic errors in
all these data sets needs 
further investigation, several, mainly phenomenological, mechanisms 
have been put forward to solve the problem of the low second peak. 
The simplest is clearly to modify standard nucleosynthesis 
so that a higher value of the baryon density parameter, 
which leads to a suppression of even
peaks becomes acceptable~\cite{Ale,ori,kapli,Lesg,hansen,dibari}. 
Another suggestion is to modify one of the 'pillars' of 
the inflationary model, the nearly scale-invariant primordial
spectrum of fluctuations by adding features on it 
\cite{kinney,louise,subir}. Also models 
with $\Omega_bh^2=0.019$ and with a 'red' tilted spectral index 
$n \sim 0.9$, even if not preferred with respect to the $\Omega_bh^2=0.03$ 
and $n=1$ models, give a reasonable $\chi^2$-fit to present 
data (see, e.g. \cite{kana,kmr,covi,moro}). Furthermore, 
in Ref.~\cite{ave,bat} it has been  found that a time-varying 
fine-structure constant can increase the compatibility
between CMB and BBN data. Finally,
a combination of inflation with topological defects which 
can contribute to the Sachs-Wolfe plateau and to the first 
peak but not to the second or third
peak, has also been proposed~\cite{Cont,Mairi} as a possible
resolution to the problem of the low secondary peaks.

Here we want to investigate whether generic defect models, 
the so-called 'causal scaling seeds'  models, can
reproduce the new data. A couple of years ago, Neil~Turok constructed a model
with scaling causal seeds which perfectly reproduced the CMB
anisotropy spectrum of inflationary models~\cite{Turokmodel}. Other
synthesized causal seed models with various heights of the acoustic
peaks are discussed in~\cite{MR,DKLS}. Spergel \& Zaldarriaga argued
that causal seeds can nevertheless be distinguished from inflationary
models by the induced polarization~\cite{SZ}. Our investigations
confirm and extend this result. But here we shall not only play with some
parameters describing the model, but we also vary cosmological
parameters, especially the total curvature which basically determines
the angular diameter distance and thereby the angular scale onto which
the peaks in the power spectrum are projected. 

\section{Unequal time correlators, seed parameters}

Let us first define the notion of 'causal scaling
seeds'. Seeds are an inhomogeneously distributed form of energy and
momentum, which provide a perturbation to the homogeneous background fluid.
In first order perturbation theory they
evolve according to the unperturbed (in general non-linear) equations
of motion. For simplicity, we assume the seeds to be coupled to the cosmic
fluid only via gravity. A counter example to this are $U(1)$ cosmic
strings. Then the resulting CMB anisotropy power spectrum, especially
the height of the first acoustic peak, depends very
sensitively on the details of the coupling of string seeds to
matter~\cite{Joao,Nathalie}. 

For uncoupled seeds the energy momentum tensor is covariantly 
conserved. To determine power spectra or other 
expectation values which are quadratic
in the cosmic perturbations, we just need to know the unequal
time correlation functions of the seed energy momentum tensor~\cite{DK,DKM},
\be
\langle \Th_{\mu\nu}(\bk,\eta) \Th^*_{\si\rho}(\bk',\eta')\rangle = 
	M^4C_{\mu\nu\si\rho}(\bk,\eta,\eta')\de(\bk-\bk')~,
\ee
where $M$ is a typical energy scale of the seeds ({\it e.g.} the
symmetry breaking scale for topological defects) which determines the
overall perturbation amplitude. Seeds are {\em causal}, if
$C_{\mu\nu\si\rho}({\bf x},\eta,\eta')$ vanishes for $|{\bf x}|> \eta+\eta'$; and
they are {\em scaling}, if $C$ depends on no other dimensional parameter
than \bk, $\eta$ and $\eta'$. Using energy momentum conservation,
statistical isotropy and
symmetries, one can then reduce $C_{\mu\nu\si\rho}(\bk,\eta,\eta')$ to five
functions of the variables $z^2=k^2\eta\eta'$ and $r=\eta'/\eta$, which are (a
consequence of causality) analytic in $z^2$~\cite{DK}. Three of these variables
describe scalar degrees of freedom, one represents vector and one
tensor contributions to the source correlator $C$. As in
Ref.~\cite{DKM}, we parameterize the scalar part by the Bardeen
potentials of the source, $\ep \equiv 4\pi GM^2=4\pi(M/M_{\rm Pl})^2$, 
\bea
\langle\Psi(\bk,\eta)\Psi^*(\bk,\eta')\rangle &=& {\ep^2\over
\sqrt{\eta\eta'}k^4}P_1(z,r)\\ 
 \langle\Phi(\bk,\eta)\Phi^*(\bk,\eta')\rangle  &=& {\ep^2\over
\sqrt{\eta\eta'}k^4}P_2(z,r)\\ 
\langle\Psi(\bk,\eta)\Phi^*(\bk,\eta')\rangle  &=& {\ep^2\over
\sqrt{\eta\eta'}k^4}P_3(z,r)~.
\eea 
The vector and tensor contributions are described by two functions
$\Si(z,r)$ and $F(z,r)$ (see Ref.~\cite{DK} from more details).


Clearly, the parameter space provided by these five functions (of two
variables)  is still enormous and it is rather impossible
 to investigate.
For a realistic
model, the parameter space is even larger due to the radiation-matter
transition which breaks scale invariance: the seed functions can be
different in the radiation and in the matter era. For global $O(N)$
defects this difference turns out not to be very important (less than
about  20\%~\cite{DKM}) it may, however, go to factors of $2$ and more
for cosmic strings~\cite{XX}.

The topological defect models studied so far, suffer from the
relatively high amplitude of vector and tensor perturbations which
contribute to the Sachs-Wolfe plateau but not to the acoustic
peaks. This is the main reasons why these models show no
significant acoustic peaks~\cite{DKM}. Here, we try to find a causal
scaling seed model which fits the CMB anisotropy data, hence vector
and tensor modes have to be suppressed. For simplicity, we set 
$\Si = F = 0$ in this study.
In this case, the sum $\Phi + \Psi$ which is due to the
anisotropic stresses in the defect energy momentum tensor is suppressed
by a factor $z^2$ on large scales, $z\ll 1$~\cite{DK}. In a first
attempt we simply set
$\Psi = -\Phi$,
which implies $P_1=P_2=-P_3 \equiv P$.

Another problem of topological defects is 'decoherence': the coupling
of different $k$-modes in the defect energy momentum tensor, which is due to
non-linear evolution,  'smears out' distinct features like peaks in
the CMB anisotropy spectrum into 
broad humps~\cite{AMFC,DKM}. To avoid this we restrict our study to so called
'perfectly coherent' models where the the unequal time correlator $P$
is simply the product of the square roots of the two corresponding
equal time correlators at $\eta$ and $\eta'$, 
\be
 P(z,r) = \sqrt{P(\sqrt{z^2r},1)P(\sqrt{z^2/r},1)}
\label{deco}
\ee
This is strictly correct if and only if the time evolution of the source is
linear.

In our numerical study described  below we investigate two families of
models.\\
{\bf Family I}\\
To enhance the acoustic peak, we use seeds which are larger in
the radiation era than in the matter era. 

\bea
 P_r(z,1) &=& {t\over 1 +(bz)^6}  \label{Pr} \\
 P_m(z,1) &=&  {1\over 1 +(bz)^6}~,  \label{Pm}  
\eea
where here the subscripts $_r$ and $_m$ indicate the radiation and
matter era respectively. The parameters $t$ and $b$ are varied to
obtain the best fit and the amplitude $\ep$ is determined by the
overall  normalization.\\
{\bf Family II}\\
The second family of models is inspired by Ref.~\cite{Turokmodel},
 which studies spherical exploding shells with $\rho+3p \propto \de(r-A\eta)$.
To formulate the model we use the source functions defined in
 Ref.~\cite{d90} which determine scalar 
perturbations of the energy momentum tensor of the seeds, $\Th_{\mu\nu}$:
\begin{eqnarray*}
 \Th_{00} &=& M^2f_\rho~,~~~  \Th_{0j} = iM^2f_vk_j
   ~, \\
  \Th_{ij} &=& M^2[f_p\de_{ij}-(k_ik_j-{k^2\over 3}\de_{ij})f_\pi]
\end{eqnarray*}
The source functions $f_{\scriptsize\bullet}$ of our models are then given by
\bean
f_\rho\! + \! 3f_p &=& {1\over \al \eta^{1/2}}{\sin( Ak\eta)\over Ak\eta}\\
f_v &=& 
{E(\eta)\over k^2\eta^{3/2}}{3\over C^2}\left[\cos
(Ck\eta) - {\sin(Ck\eta)\over Ck\eta} \right]
\eean
with $\al =(\dot a/a)\eta$ and $E=(4-2/\al)/(3-12\al)$.
The functions $f_\rho$ and $f_\pi$ are then determined by energy
momentum conservation~\cite{d90},
\bean
\dot f_\rho + k^2f_v +{\al\over \eta}(f_\rho +3f_p) &=& 0 \\
\dot f_v + 2{\al\over \eta}f_v -f_p +{2\over 3}k^2f_\pi &=& 0 ~.
\eean
The function $E$ is chosen such that the power spectrum of $f_\pi$ is
white noise on super horizon scales, a condition which is required
for purely scalar causal seeds~\cite{DK}.
This leads to the Bardeen potentials~\cite{d90}
\bea
 \Phi &=& {\ep\over k^2}(f_\rho + 3{\al\over \eta}f_v)~,
 \label{phT}\\
 \Psi &=& -\Phi-2\ep f_\pi~.   \label{psT}
\eea
Here the seed functions are actually not given as random variables
but as square-roots of power spectra, and one has always to keep in
mind that we assume perfect coherence. Of course one can also regard
Eqs.~(\ref{phT},\ref{psT}) as mere definitions with
\bea
 P_1(z,1) &=& \eta k^4(\Psi)^2/\ep^2 ~, \\
 P_2(z,1) &=&  \eta k^4(\Phi)^2/\ep^2 ~, \\
 P_3(z,1) &=&   \eta k^4\Psi\Phi/\ep^2 = -\sqrt{ P_1(z,1) P_2(z,1)} ~.
\eea
With a somewhat lengthy calculation one can verify that $E$ is
chosen such that $f_\pi \propto$ const. for $z\ll 1$ and the
functions $P_i(z,1)$ are analytic in $z^2=(k\eta)^2$. This
family of models is described by the parameters $A$ and $C$, which have to
satisfy $0 < A,~ C \le 1$ for causality.
Also here  one can choose different amplitudes for the
source functions in the radiation and matter era by introduction of
the additional parameter $t\neq 1$.

\section{Angular diameter distance, cosmological parameters}

Seeds generically produce isocurvature perturbations.
These models, for a flat universe, predict a position of the first
peak at $\ell \sim 350$, which is definitely incompatible
with the recent CMB observations (\cite{juan}, \cite{enqvist}).
However, the tight constraints on the flatness of the universe 
obtained from CMB data analysis are based on the assumption of 
adiabatic primordial fluctuations. 
Using this loophole, it is possible to construct 
 closed $\Lambda$-dominated isocurvature models which have the first
acoustic peak in the observed position.

For a given seed-model, the position of the first acoustic peak is
determined primarily  by the
angle subtended by the acoustic horizon $\la_{ac}$ at decoupling time,
$\eta_{dec}$. 
The angle under which a given comoving scale $\la$ at conformal time
$\eta_{dec}$ is seen on the sky is given by
$  \theta(\la) =\la/\chi(\eta_0-\eta_{dec})$,
where
\[ \chi(y) =\left\{\begin{array}{ll}
   \sin(y) & \mbox{ if }~~ K>0 \\
   \sinh(y) & \mbox{ if }~~ K<0 \\
   y& \mbox{ if }~~ K =0~. \end{array}\right.
\]
(K denotes the curvature of 3-space.)\\
As the harmonic number $\ell$ is inversely proportional to the angle
$\theta$, this yields
$\ell_{\rm peak} \simeq R \ell_{\rm peak}^{\rm flat}$ ~where~ 
$R =\theta_{ac}^{\rm flat}/\theta_{ac}$~.
The well-known expressions for the conformal time (see {\em e.g.}
Ref.~\cite{Pee}) $\eta_{\rm dec}$ and $\eta_0$ are
\begin{eqnarray*}
\eta_{\rm dec} &=&2\sqrt{|\Omega_K|} \over \Omega_m
	\sqrt{\Omega_{\rm rad} +\Omega_m/(z_{\rm dec}+1)}  \\
\eta_0-\!\eta_{\rm dec} 
 &=&\sqrt{|\Omega_K|}\!\int_0^{z_{\rm dec}}\!\!
	{dz\over [\Omega_m(1\!+\! z)^3 + \Omega_K(1\!+\! z)^2
	+\Omega_\Lambda]^{1/2}}
\end{eqnarray*}
 which leads to
\bean \theta_{ac}^{\rm flat} &\equiv& 
   \theta_{ac}(\Omega_m=1,\Omega_{\Lambda}=0, \Omega_K=0) \\
  &=& c_s\eta_{\rm dec}/(\eta_0-\eta_{\rm dec})\\
  &=& c_s\sqrt{\Omega_{\rm rad} +1/(z_{\rm dec}+1)} ~,
\eean
where $c_s =1/\sqrt{3(1+3\Om_b/4\Om_{\rm rad}(1+z_{\rm dec}))}$ denotes the
adiabatic sound speed of the baryon/photon plasma at decoupling.
We then find
\be R=\frac{1}{2}
\frac{\Omega_m \sqrt{\Omega_{\rm rad}+1/(z_{dec}+1)}}{
\sqrt{|\Omega_K|} \sqrt{\Omega_{\rm rad}+\Omega_m/(z_{dec}+1)}}
\chi(\eta_0-\!\eta_{\rm dec}) ~.
\ee
 Neglecting $\Omega_{\rm rad}$ this reduces to the result of Ref.~\cite{BE}
(the factor $1/2$ is missing in their formula),
\[ 
R=\frac{1}{2} \sqrt{\frac{\Omega_m}{|\Omega_K|}}\chi(\eta_0-\!\eta_{\rm dec}).
\] 
An interesting point is that for
$\Omega_m \ra 0$ the quantity $R$ depends very sensitively on
$\Omega_{\Lambda}$. Thus, we can have important shifts in
the power spectrum, $R \sim 0.6$ say,  with relatively
small deviations from flatness ($\Omega_m=0.3$,
$\Omega_{\Lambda}=0.9$, $\Omega_K=-0.2$).
In Ref.~\cite{BE} the authors have shown
that the simple prescription $\ell \ra R\ell$ reproduces the CMB power
spectra for curved universes within a few percent.
On lines of constant $R$,  CMB power spectra are nearly degenerate.  
In this study we use this simple prescription to
rescale the flat spectrum. Thereby, we make sure that the value of
$\Om_m$ used in the spectrum calculation agrees roughly with the value
preferred by our best fit value of $R$ and the super-novae
constraint~\cite{SNIa}, which can be cast in the form $\Om_m \simeq
0.75\Om_\Lambda-0.25$. $\Om_m$ determines the time of equal matter
and radiation and thus influences the early integrated Sachs-Wolfe
effect, which contributes to the spectrum right in the region of the
first peak. We therefore get a better approximation if we use
the correct value for $\Om_m$.

\section{Results}
To analyze family~I given by Eqs.~(\ref{Pr},\ref{Pm}), we have
investigated a grid of models in $(t,b)$ space with $1<t<2$ and 
$0.1<b<1$. To make sure that the models are causal, we Fourier
transform the correlation function into real space, cut it
at $|\bf x|=\eta+\eta'$ and transform it back. This procedure prevents
acausal early decay of the correlation function; we find that
models with $b>1$ do not significantly differ from $b=1$ after
application of this causality constraint. 

For each model in our grid we then search the values $R$ and the
normalization $\ep$ which minimize $\chi^2$ when compared with
the B98~\cite{B98} and Maxima~\cite{MAX} data. We also allow for an
overall re-calibration of the B98 data by 20\% and of the Maxima
data by 8\%. In Figs.~1 and 2 we show the temperature and polarization
spectrum for the best model (long dashed lines). This model
corresponds to the best fit 
parameters $t=2.2,~b=1/9,~\Om_m=0.35$ and $R=0.53$. It has a value of
$\chi^2 =38$, which, for $22$ points and $4$ parameters ($t$, $b$, $R$
and the normalization), 
it is excluded at more than $99 \%$ c.l. if Gaussian statistics is assumed.
The main disagreement, also for this model, is from the high second peak.
Assuming $\Omega_bh^2=0.03$ brings the model in better agreement with
$\chi^2=31$.
However, as is clearly visible from Fig.~\ref{ct}, another main contribution to
$\chi^2$ comes from the last two maxima points. If these points are
disregarded, the model has a $\chi^2$ which is somewhat lower than the
one of a typical $\La$CDM model (short dashed line). But it is clearly
visible that shifting the spectrum does not only move the peak into
the correct position but it also reduces the width of the peak which
is already a problem for this model.
It is conceivable that the introduction of a small amount of
decoherence into the model might somewhat enlarge the peak width and
lead to a better fit. Nevertheless, present data already does not favor
coherent closed isocurvature models over the corresponding flat adiabatic 
models. Furthermore, since the model is
closed, $\Om_\La+\Om_m \sim 1.2$, the secondary peaks are at smaller
values of $\ell$ than in a flat model, which makes this model easily
distinguishable from a flat model with sufficiently accurate
measurements as envisaged by the Planck satellite~\cite{Planck}. This
difference of the inter-peak distance which is
given only by the values of cosmological parameters like
$\Om_\La+\Om_m$, is also present in the the polarization
spectra (see Fig~\ref{cp}). Another important difference is that,
in general, the signal in the $50 \le \ell \le 500$ band is 
$\sim 50 \%$ higher for the isocurvature model. 
CMB polarization is produced by Thomson scattering which is 
active only on sub-horizon scales: at fixed $\ell$, 
the relevant physical scales are  more inside the horizon in the
closed model and so the contribution to the signal is higher.

\begin{figure}[ht]
\centerline{\psfig{figure=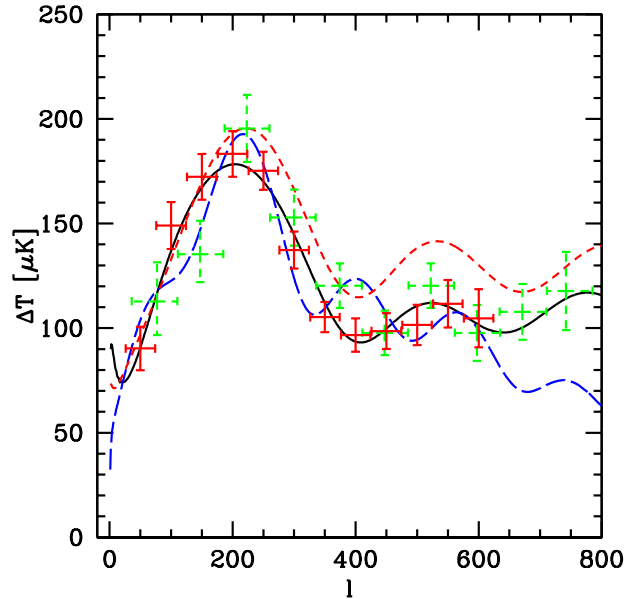,width=8.3cm}}
\caption{The CMB temperature anisotropy spectrum
$\ell(\ell+1)C_\ell^{(T)}$ for our best fit model of family I (long
dashed, blue) and family II (solid, black) is compared with the B98 and
the Maxima-1 (short dashed, red) data. The family I model is
a rather good fit to the first peak, even if is a closed model
 ($\Omega \sim 1.2$). The family II model is flat and is in perfect 
 agreement with the data ($\chi^2=14./18$) even with $\Omega_bh^2=0.019$, 
 as BBN constraints suggest.
A standard inflationary spectrum with
$h=0.65,~ h^2\Om_b=0.019,~ \Om_{cdm}=0.3,~ \Om_\La=1-\Om_m$
is also indicated (short dashed). \label{ct}}
\end{figure}

\begin{figure}[ht]
\centerline{\psfig{figure=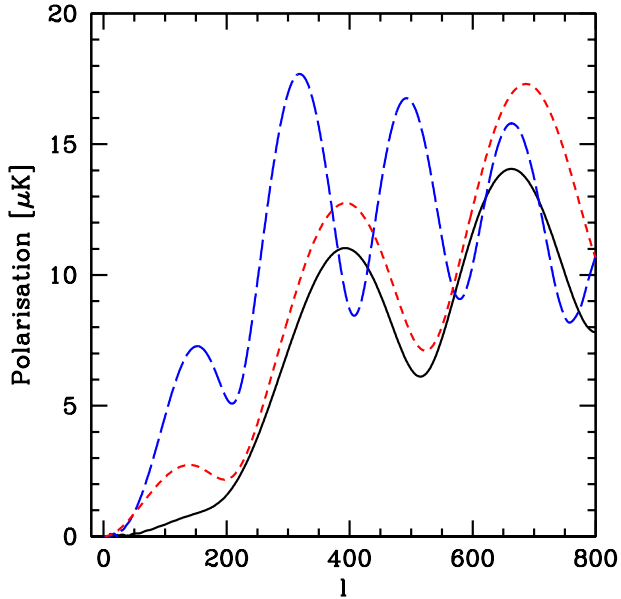,width=8.3cm}}
\caption{The CMB polarization  spectrum $C_\ell^{(P)}$s  for our best
fit model of family I  (long dashed, blue)
and family II (solid, black) is compared with a standard
inflationary spectrum with the same parameters as above (short dashed,
red). The family I model predicts a larger r.m.s. polarization signal
in the band $50 \le \ell \le 500$.
On the contrary, the lack of intermediate scale polarization 
at $\ell \le 200$ in the family II model is clearly visible.
 \label{cp}}
\end{figure}

A much better fit can be achieved by the models of family~II.
To study these models we have varied $0.3\le A,C\le 1$,
and $0.5\le t\le 1.5$. Our
best fit model with a value of $\chi^2=14.5$ for $22$ points and $5$
parameters ($A,~C,~t,~R$ and the normalization) is in very good agreement with
the data (see fig.~\ref{ct}, solid line), and, up to the second peak,
is actually quite similar to a model with high baryon content.  
The model shown corresponds to the best fit
parameters $A=1,~C=0.85,~t=0.8,$ and $R=1$.
In this model which is flat and causal, the first
peak in the polarization spectrum is suppressed, as
has been noted in Ref.~\cite{SZ}  (see fig.~\ref{cp}, solid line). 

%

\section{Conclusions}

In this paper we have shown that causal scaling seed models for
structure formation can
reproduce the recent CMB anisotropy data~\cite{B98,MAX}. A first very simple
closed model (family I) can be brought in reasonable agreement with all
but the last two Maxima-1 points for a baryon density which is
not compatible with the nucleosynthesis constraint. 
It is interesting to note that present data already slightly  disfavors
closed isocurvature models since they have a peak which is narrower
than what is preferred by the data. 
A somewhat more refined model  (family II) is, for a
suitable choice of parameters, in excellent agreement with all data
points in a flat, $\La$-dominated universe. The cosmological
parameters of our best fit model~I are
$\Om_m=\Om_{cdm}+\Om_b=0.35,~\Omega_\La=0.85, h=0.65,~h^2\Om_b=0.019$ and
those of model~II are 
$\Om_m=\Om_{cdm}+\Om_b=0.4,~\Omega_\La=0.6, h=0.65,~h^2\Om_b=0.019$. This
model is preferred by the data with respect to the 'concordance model' with 
$\Om_m=0.3,~\Omega_\La=0.7,~h=0.65,~h^2\Om_b=0.019$ and
inflationary initial conditions. These models can, however, be clearly
distinguished from inflationary models by future experiments either
measuring the secondary peaks or the polarization spectrum. The first
one, a closed model,  has smaller inter-peak distances than flat
inflationary models  (see fig.~\ref{ct}), a definitive
lower amplitude of temperature fluctuations for $\ell \ge 650$ and 
a {\it greater} r.m.s. amplitude of polarization for $50 \le \ell \le 550$;
 in the second one the first peak $\ell \sim 150$ in the
polarization spectrum  is not present (see fig.~\ref{cp}), which is a
consequence of causality, and the polarization 
amplitude is generally lower in the band $0 \le \ell \le 800$.

To achieve this agreement we have suppressed vector and tensor
perturbations and have assumed perfectly coherent fluctuations. We
believe that it is quite improbable that topological defects 
from a GUT phase transition have such a behavior. Nevertheless, there might
be some other scale-invariant causal physical mechanism (e.g. some
spherically symmetric 'neutrino explosions', see
Ref.~\cite{Turokmodel}) leading to 
seeds of this or similar type. Clearly, we only have a satisfactory
model of structure formation if also the physical origin of the
'seeds' is clarified. However, the point of this work 
was not to find ``the correct model of large scale structure formation'' 
but mainly to investigate, in a phenomenological but at the same
time physically motivated way, to what extent the present values of
the cosmological parameters derived from accurate CMB data analysis
can still be plagued by the assumption of the underlying theoretical model.
We have seen {\em e.g.} that flatness, $\Om_m+\Om_\La=1$ is not mainly
supported by the position of the first peak but by its {\em width}. Clearly,
once secondary peaks are unambiguously detected, the inter peak
distance will represent another direct measure of the total curvature. 

This investigation is rather important especially if some of 
the parameters obtained assuming the standard inflationary
model are in significant disagreement with complementary, more direct
observations, as the high $\Omega_bh^2$ value seems to suggest.
While present CMB data can be regarded as a triumph 
for a scenario based on primordial adiabatic fluctuations, 
we have presented here phenomenological models, 
based on isocurvature fluctuations, that also give a good fit 
to the CMB data.
Fortunately, the concrete models proposed here have peculiar 
characteristics that future CMB experiments will be able to
detect\footnote{A systematic study of the precision with which
cosmological parameters can be determined by the Planck
satellite if the model space is enlarged to allow for arbitrary 
isocurvature models {\em without seeds}, is presented 
in Ref.~\cite{Bucher}}.
\vspace{0.2cm}\\
{\bf Acknowledgments}\hspace{0.5cm} We thank Louise Griffith,
Mairi Sakellariadou, Jean-Philippe Uzan and Julien Devriendt 
for stimulating discussions.\\
This work is supported by the Swiss National Science Foundation.

}

\begin{thebibliography}{99}
\bibitem{h} H.R. Harrison, { Phys. Rev.} {\bf D1}, 2726 (1970).
\bibitem{z} Y.B. Zel'dovich, { Mon. Not. Roy. Ast. Soc.} {\bf160}, 1
	(1972).
\bibitem{SPT} U. Pen, D. Spergel \& N. Turok, { Phys. Rev.} {\bf
	D49}, 692 (1994).
\bibitem{DZ} R. Durrer \& Z. Zhou, { Phys. Rev.} {\bf D53}, 5394 (1996).
\bibitem{Shellard} B. Allen at al.,
	Phys. Rev. Lett. {\bf 77}, 3061 (1996);   preprint  {\tt
	astro-ph/9609038} .  
\bibitem{Roma}R. Durrer, in {\em 
	Topological Defects in Cosmology}, eds. M. Signore \&
	F. Melchiorri, World Scientific (1998); preprint
	               {\tt astro-ph/9703001}.
\bibitem{DGS} R. Durrer, A. Gangui \& M. Sakellariadou, {\em
	Phys. Rev. Lett.} {\bf 76}, 579 (1996).
\bibitem{Seljak}  U. Pen, U. Seljak \& N. Turok,
	Phys. Rev. Lett. {\bf 79} 1611 (1997); 
	preprint {\tt astro-ph/97004165 }.
\bibitem{DKM}R. Durrer, M. Kunz \& A. Melchiorri,  Phys. Rev {\bf
	D59} 123005 (1999); preprint
	{\tt astro-ph/9811174}.
\bibitem{Albrecht}R. Battye, J.Robinson \& A. Albrecht,
	Phys. Rev. Lett. {\bf 80}, 4847 (1998);
	preprint {\tt astro-ph/9711336}.
\bibitem{Joao}C. Contaldi, J. Magueijo \& M. Hindmarsh,
	Phys. Rev. Lett. {\bf 82}, 679 (1999); preprint {\tt astro-ph/9808201}.
\bibitem{Vachaspati}L. Pogosian \& T. Vachaspati, Phys. Rev. {\bf
	D60},  083504 (1999); preprint {\tt astro-ph/9903361}.
\bibitem{toco} A. Miller, {\em et al.}, Astrophys.J. 524 (1999) L1-L4.
\bibitem{B97} P. Mauskopf, {\em et al.}, ApJ, {\bf 536}, L59; 
preprint {\tt astro-ph/991144} (1999). A. Melchiorri, {\em et al},
ApJ, {\bf 536}, L63; preprint {\tt astro-ph/991145} (1999).
\bibitem{B98}P. DeBernardis {\em et al.}, Nature {\bf 404}, 955; 
preprint {\tt astro-ph/0004404} (2000).
\bibitem{MAX}S. Hanany {\em et al.}, accepted for publication in Astrophys. J. Letters; preprint {\tt astro-ph/0005123} (2000).
\bibitem{pierpol} M. White, D. Scott, E. Pierpaoli, Astrophys. J.;
preprint {\tt astro-ph/0004385} (2000).
\bibitem{tegman} M. Tegmark, M. Zaldarriaga, Phys. Rev. Lett., 85, 2240 (2000);
preprint {\tt astro-ph/0004393}.
\bibitem{tegman2} M. Tegmark, M. Zaldarriaga, A. Hamilton, submitted 
to ApJ; preprint {\tt 0008167} (2000).
\bibitem{lange} A. Lange et al, PRD in press; preprint {\tt astro-ph/0005004}
(2000).
\bibitem{jaffe}A. Jaffe {\em et al}, Phys. Rev. Lett., in press; preprint {\tt astro-ph/0007333} (2000).
\bibitem{burles}S. Burles, K. M. Nollet, M. S. Turner, preprint
	{\tt astro-ph/0010171} (2000).
\bibitem{Ale}S. Esposito, G. Mangano, A. Melchiorri,
	G. Miele, O. Pisanti,  PRD, in press; preprint
	{\tt astro-ph/0007419} (2000).
\bibitem{ori} M. Orito et al., (2000); preprint {\tt astro-ph/0005446}.
\bibitem{kapli}M. Kaplinghat, M. S. Turner (2000); preprint {\tt astro-ph/0007454}. 
\bibitem{Lesg}J. Lesgourgues \& M. Peloso, 
Phys. Rev. {\bf D62}, 81301 (2000); preprint {\tt astro-ph/004412}.
\bibitem{hansen}S. Hansen, F. Villante, Phys.Lett. B486, 1-5, (2000);  
	preprint {\tt astro-ph/0005114}.
\bibitem{dibari}P. Di Bari and R. Foot, preprint {\tt hep-ph/0008258} (2000).
\bibitem{kinney}W. H. Kinney, (2000); preprint {\tt astro-ph/0005410}.
\bibitem{louise}L. Griffiths, J. Silk, S. Zaroubi,
	preprint {\tt astro-ph/0010571} (2000).
\bibitem{subir} J. Barriga, E. Gazta\~{n}aga, M. Santos and S. Sarkar, 
	in preparation (2000).
\bibitem{kana} T. Kanazawa, M. Kawasaki, N. Sugiyama, T. Yanagida,
	preprint {\tt astro-ph/0006445} (2000).
\bibitem{kmr} W. H. Kinney, A. Melchiorri, A. Riotto; PRD in press;
	preprint {\tt astro-ph/0007375} (2000).
\bibitem{covi} L. Covi, D. Lyth, preprint {\tt astro-ph/0008165} (2000).
\bibitem{moro} T. Moroi, T. Takahashi, preprint {\tt 0010197} (2000).
\bibitem{ave} P.P. Avelino,  C.J.A.P. Martins, G. Rocha and P. Viana, 
	preprint {\tt astro-ph/0008446}, (2000).
\bibitem{bat} R. A. Battye, R. Crittenden, J. Weller, preprint 
	{\tt astro-ph/0008265}, (2000).  
\bibitem{Cont}C. Contaldi, preprint
	{\tt astro-ph/0005115} (2000).
\bibitem{Mairi}F. Bouchet, P. Peter, A. Riazuelo \& M. Sakellariadou,
	preprint {\tt astro-ph/0005022} (2000).
\bibitem{Turokmodel}N. Turok, Phys. Rev. Lett. {\bf 77}, 4138 (1996).
\bibitem{MR}R. Durrer \& M. Sakellariadou, {
	Phys. Rev.} {\bf D56}, 4480 (1997).
\bibitem{DKLS}R. Durrer, M. Kunz, C. Lineweaver \&  M. Sakellariadou,
	Phys. Rev. Lett. {\bf 79}, 5198 (1997).
\bibitem{SZ}D. Spergel \&  M. Zaldarriaga, Phys. Rev. Lett. {\bf
	79}, 2180 (1997).
\bibitem{DK}R. Durrer \&  M. Kunz, { Phys. Rev.} {\bf D57}, R3199
	(1998).
\bibitem{Nathalie}A. Riazuelo, N. Deruelle \& P. Peter Phys.Rev. {\bf
	D61}, 123504 (2000). 
\bibitem{XX}P.   Shellard, private communication.
\bibitem{AMFC}J. Magueijo, A. Albrecht, P. Ferreira \& D. Coulson
	Phys. Rev. {\bf D54}, 3727 (1996); preprint {\tt astro-ph/9605047}.
\bibitem{d90}R. Durrer, { Phys. Rev.} {\bf D42}, 2533 (1990).
\bibitem{juan} E. Pierpaoli, J. Garcia-Bellido, \& Stefano Borgani,
        preprint  {\tt  hep-ph/99009420}, (1999).
\bibitem{enqvist}K. Enqvist, H. Kurki-Suonio and J. Valiviita;
	Phys.Rev. D62, 103003, (2000).
\bibitem{Pee}P.J.E. Peebles, Principles of Physical Cosmology,
	Princeton University Press (1993).
\bibitem{BE}R. Bond \& G. Efstathiou,  preprint
	 {\tt astro-ph/9807103} (1998).
\bibitem{SNIa}A. Riess  {\em et al.} AJ, {\bf 116}, 1009 (1998);
     S. Perlmutter {\em et al.}, Astrophys. J. {\bf 517}, 565 (1999).
\bibitem{Planck} Information on the Planck satellite mission can be found on
	its home page: \\  {\tt
	http://astro.estec.esa.nl/SA-general/Projects/Planck/}. 
\bibitem{Bucher} M. Bucher, K. Moodley \& N. Turok, preprint  {\tt
	astro-ph/0007360 }, (2000).
\end{thebibliography}
\end{document}